\documentclass[aip,twocolumn,amssymb,amsfonts,amsmath,%
reprint,superscriptaddress,letterpaper,showpacs,floatfix]{revtex4-1}
\usepackage{graphicx}
\usepackage{dcolumn}
\usepackage{bm}
\usepackage{mathtools}
\usepackage{mathrsfs}
\usepackage{color}
\newcommand{\imsize}{\columnwidth}
\begin{document}
\title[]{Electron correlation in solids via density embedding theory}
\author{Ireneusz W. Bulik}
 \affiliation{Department of Chemistry, Rice University, Houston, Texas 77005, USA}
\author{Weibing Chen}
 \affiliation{Department of Chemistry, Rice University, Houston, Texas 77005, USA}
\author{Gustavo E. Scuseria}
 \affiliation{Department of Chemistry, Rice University, Houston, Texas 77005, USA}
 \affiliation{Department of Physics and Astronomy, Rice University, Houston, Texas 77005, USA}
\date{\today}
\begin{abstract}
Density matrix embedding theory (Phys. Rev. Lett. {\bf 109}, 186404 (2012))
and density embedding theory (Phys. Rev. {\bf B 89}, 035140 (2014)) have recently been
introduced for model lattice Hamiltonians and molecular systems. In the present
work, the formalism is extended to the {\it ab initio} description of infinite
systems. An appropriate definition of the impurity Hamiltonian for such systems
is presented and demonstrated in cases of 1, 2 and 3 dimensions, using coupled cluster
theory as the impurity solver. Additionally, we discuss the challenges related to
disentanglement of fragment and bath states. The current approach yields results
comparable to coupled cluster calculations of infinite systems even when using
a single unit cell as the fragment. The theory is formulated in the basis of Wannier
functions but it does not require separate localization of unoccupied bands.
The embedding scheme presented here is a
promising way of employing highly accurate electronic structure methods for
extended systems at a fraction of their original computational cost.
\end{abstract}

\maketitle
\section{Introduction}

Electron correlation plays a crucial role in understanding most physical
phenomena in molecules and extended systems.  While highly
accurate many-body approaches can be nowadays routinely employed in molecular systems,
solid state applications remains dominated by density functional theory (DFT).
Despite the undeniable success of DFT in extended systems,
\cite{ParrYang,JCP.2014.140,JCP.2012.136} there are significant limitations appearing
from the approximate form of the exchange-correlation functional
\cite{Science.298.2002} and much work remains to be done in order to ensure
systematically improvable predictions.
\cite{Science.321.2008,PRL.100.136406.2008,PRB.83.035119.2011,
JP.CM.2013.43.435503,PRB.87.035117.2013,JCP.2004.121.1187,JCP.2005.123.174101,PRB.83.205128,
JP.CM.2012.14.145504,PRB.87.035107} An alternative route for incorporating
electron correlations is to employ wavefunction-based methods. Recently,
significant progress in applying such many-body theories for solid state
problems has been made.
\cite{JCP.2008.129.204104,PhysRevB.80.085118,
JCP.132.2010.151101,Nature.2013.493.365,JCTC.2014.10.1698}

Size-extensive, wavefunction-based approaches to solids treat the system
as a whole, imposing translational symmetry and Brillouin zone integration.
Finite order perturbation theory
\cite{CP.178.1993,PRB.50.14791,PRB.51.16553,
JCP.1996.104.8553,JCP.106.1997.5554,JCP.109.1998,
JCP.2001.115.9698,JCP.130.2009.184103,JCP.133.2010.0741067}
and coupled cluster (CC) methods
\cite{JCP.106.1997.10248, TCA.104.2000.426,JCP.120.2004.2581,RodsBook}
have been formulated and implemented for infinite systems.
Alternatively, the numerical complexity associated
with numerous electronic degrees of freedom in solids has been
simplified, for example, by means of the method of increments.
\cite{PRB.46.6700.1992,CPL.1992.548,JCP.1992.97.8449,CP.1997.224.121,PhysRep.2006.428.1}

In the present work, electron correlation in extended systems is accounted for via an
embedding approach. The infinite periodic problem is transformed into one
of a small fragment (unit cell) entangled with an effective bath.  In order to
define the bath, we employ the recently introduced density embedding theory (DET),
\cite{PhysRevB.89.035140} which is a simplification of density matrix
embedding theory (DMET).  \cite{PhysRevLett.109.186404,JCTC.9.1428}
Here, an approximate solution to the infinite periodic system, typically
Hartree-Fock, is used to construct two basis sets. The first basis is associated
with a small part of the lattice (fragment) whereas the second is used to
describe the excluded complement (bath). Subsequently, one solves the
many-body problem for the fragment plus bath, a so-called impurity problem.  In
this way, correlations between the fragment and the rest of the system are
represented by a many-electron environment, not a single-particle potential.
\cite{JCTC.9.1428} The Hartree-Fock (HF) choice for an approximate solution of the
infinite solid affords the desired bases in a trivial algebraic manner via
diagonalization. Moreover, the resulting Hamiltonian describing the
fragment-bath interaction is defined in a much smaller single-particle Hilbert
space as compared to the full lattice Hamiltonian. Thus, this construction opens
the possibility of employing highly accurate many-body techniques to tackle
extended systems.

Indeed, DMET and DET with  exact
diagonalization\cite{PhysRevLett.109.186404,JCTC.9.1428,Booth2013,PhysRevB.89.035140}
and density matrix renormalization group\cite{PRB.89.165134} as
impurity solvers have been shown to provide high-quality
descriptions of model Hamiltonians and molecular systems. In the
present work we extend the applicability of this novel
approximation to realistic extended systems. Such problems seem a
promising niche for this embedding scheme.  The definition of
fragment is naturally dictated by the periodicity of the system
which greatly simplifies the procedure.  We propose a way of
defining the local Hamiltonian that guarantees exactness of
mean-field in mean-field embedding and facilitates coping with the
Coulomb problem in solids. Additionally, we discuss practical
issues related to preparing the fragment and bath basis for an
infinite number of electrons and the problem of disentangled
states.

While in the current work we benchmark the DET scheme by
describing the fragment-bath interaction at the coupled cluster
level of theory, we stress that the methodology is flexible enough
to accommodate any correlated wavefunction method. Indeed, the
purpose of DET is to provide a finite, small dimension effective
Hamiltonian which should account for locally important degrees of
freedom. Then, any many-body technique may be employed to solve
the impurity problem at hand. In particular, very accurate and
computationally affordable schemes
\cite{JCP.2013.139,PRB.87.235129} may be used to study extended
systems with realistic unit cell sizes.
\section{Theory and formalism}
In order to keep this paper self-contained, we present in this
section a basic introduction to density matrix and density
embedding theories. More details can be found in the original
papers.
\cite{PhysRevLett.109.186404,JCTC.9.1428,Booth2013,PhysRevB.89.035140,PRB.89.165134}
Subsequently, we discuss the changes required to apply the
formalism to the {\it ab initio } Hamiltonian of extended systems.
This includes the Schmidt decomposition of Slater determinants for
infinite systems and the formulation of an impurity Hamiltonian
that allows us to deal with the Coulomb divergence in solids.
Unless otherwise specified, single-particle indices denote both
spin and spatial coordinates. In the context of periodic systems,
cell coordinates are implicitly included except when confusion may
arise.

\subsection{Mean-field based embedding theories}
Density matrix embedding theory and its simplification, density
embedding theory, are projections of the exact Hamiltonian onto a
basis obtained by Schmidt decomposition of the ground state
wavefunction $|\Psi\rangle$. To be precise, one may cast
$|\Psi\rangle$ into \cite{PhysRevLett.109.186404}
\begin{align}
|\Psi\rangle = \sum_i \lambda_i |\alpha_i\rangle |\beta_i\rangle ,
\end{align}
where $|\alpha\rangle$ represents the part of the system of interest, the fragment,
whereas $|\beta\rangle$ represents the rest of the system, the bath. With such states at hand,
an impurity Hamiltonian is defined,
\begin{align}
\hat{H}_{\textrm{imp}} = \sum_{ijkl}
|\alpha_i\rangle |\beta_j\rangle \langle \alpha_i | \langle \beta_j |
\hat{H}
|\alpha_k \rangle | \beta_l \rangle \langle \alpha_k | \langle \beta_l | .
\end{align}
which has the same ground state as the exact Hamiltonian.\cite{PhysRevLett.109.186404}
The fragment and bath basis states are, in principle, many-electron states.

In order to make calculations practical, DMET and DET
replace the exact ground state with a mean-field approximation,
i.e. $|\Psi\rangle$ is replaced with $|\Phi\rangle = \Pi_{p} a{}^\dagger_p |0\rangle$,
where $a{}^\dagger_p$ creates a hole state $|\phi_p\rangle$ and $|0\rangle$ is the bare vacuum.
The hole creation operators are obtained from a mean-field approximation,
here the Hartree-Fock transformation $\mathbb{D}$ of bare fermion
operators $c^\dagger$:
\begin{align}
a{}^\dagger_p = \sum_{\mu} \mathbb{D}_{\mu p} c{}^\dagger_{\mu}.
\end{align}
The mean-field solution is obtained
for the Hamiltonian of interest augmented with an effective one-body potential
$v$,
\begin{align}
\label{FullHam}
\hat{H} = \sum_{\mu \nu} h_{\mu \nu} c{}^\dagger_\mu c_\nu +
\frac{1}{4} \sum_{\mu \nu \lambda \sigma} V_{\mu \nu \lambda \sigma}
c{}^\dagger_\mu c{}^\dagger_\nu c_\sigma c_\lambda
+ \sum_{\mu \nu} v_{\mu \nu} c{}^\dagger_\mu c_\nu .
\end{align}
The meaning of this potential will soon become apparent.

With the above approximation, the task of performing the Schmidt decomposition of
the wavefunction describing the whole system amounts to a rather trivial
algebraic problem \cite{JPA.39.L85}. Defining single
particle basis associated with a chosen subsystem of the entire
problem, $|F\rangle$, one constructs a projection operator onto the fragment
$\hat{P}_F = \sum_{i} |F_i\rangle \langle F_i|$ and its complement $\hat{P}_B =
\hat{\mathbb{I}} - \hat{P}_F$. The latter projects onto the bath states. With such
tools at hand, one may construct an overlap matrix $\mathbb{M}$,
\begin{align}
\label{MMat}
\mathbb{M}_{pq} = \langle \phi_q | \hat{P}_F | \phi_p \rangle ,
\end{align}
where indices $p$ and $q$ denote the hole states.
Diagonalizing the above overlap matrix $\mathbb{V} d \mathbb{V}^\dagger = \mathbb{M}$
yields at most min($n_e$,$n_F$) eigenvalues $d_i$ different from zero
where $n_e$ and $n_F$ denote the number of electrons in the system and
the number of single-particle states associated with the fragment, respectively.
The eigenvectors corresponding to such eigenvalues are then normalized
to construct fragment ($|f\rangle$) and bath ($|b\rangle$) states
\begin{align}
\label{ConFrag}
|f_i\rangle &= \sum_{p} \frac{\mathbb{V}{}^\star_{p i}}{\sqrt{d_i}} \hat{P}_F |\phi_p\rangle  \nonumber \\
|b_i\rangle &= \sum_{p} \frac{\mathbb{V}{}^\star_{p i}}{\sqrt{1-d_i}} \hat{P}_B |\phi_p\rangle .
\end{align}
The states that correspond to zero eigenvalue are called {\it cores} and
are discarded from consideration in the impurity problem.

In practical applications, one would like to retain all the fragment states.
This requires special care when the eigenvalues $d$ are close to 1 or
0.  Such complications are addressed in Appendix \ref{LinDep}. Right now,
let us stress the consequences of the above approximation which is a key step
in the present work. The most prominent result is that the Schmidt decomposition
yields single-particle bases that can be further employed in the construction of
the impurity Hamiltonian.  This fact greatly facilitates computations.
Additionally, the number of entangled states (or equivalently the number of
non-zero eigenvalues $d$) is defined by the fragment single-particle basis. If
one chooses the fragment to be small, accurate many-body techniques can be
applied to solve the impurity problem, which we now proceed to introduce.

Let us define creation operators in the impurity basis: $f^{\dagger}$,
$b^{\dagger}$ and $e^\dagger$; $f^{\dagger}$ and $b^{\dagger}$ are
associated with fragment and bath states, respectively. We will use $e^\dagger$ for general
states which can be either fragment or bath.  The impurity problem is defined
by the Hamiltonian,
\begin{align}
\hat{H}_{\textrm{imp}} &= \sum_{e e^\prime} \tilde{h}_{e e^\prime} e{}^\dagger e^\prime +
\frac{1}{4}\sum_{e e^\prime e^{\prime\prime} e^{\prime\prime\prime}}
\tilde{V}_{e e^\prime e^{\prime\prime} e^{\prime\prime\prime}}
e{}^\dagger e{}^{\prime\dagger} e^{\prime\prime\prime} e^{\prime\prime}  \nonumber \\
&+
\sum_{b b^\prime} \tilde{v}_{b b^\prime} b{}^\dagger b^\prime ,
\end{align}
where $\tilde{h}$ and $\tilde{V}$ are one- and (antisymmetrized) two-body
terms of the Hamiltonian projected onto the embedding basis.
The additional potential $\tilde{v}$, acting only in the bath subspace of
the impurity, is introduced to enforce a suitable chosen convergence criterion.
In the present work, we make a diagonal ansatz for the effective potential
($\tilde{v}_{ij} = v \delta_{ij}$ hence
$v_{\mu \nu} = v \delta_{\mu \nu}$ in Eq. \ref{FullHam}),
which corresponds to a chemical potential in the bath.
We find such a potential by requiring
the proper number of electrons in the fragment, on average. Let us stress
that, for periodic systems, the average number of electrons per
unit cell is known and well defined.  The reader is referred to Ref.
\onlinecite{PhysRevB.89.035140} for other possible choices of the effective
potential.

At this point, let us make a few remarks concerning the meaning of the impurity
Hamiltonian. Assuming that $\mathbb{M}$ contains $n_F$ eigenvalues different
from 1 or 0, this Hamiltonian describes a system of $n_F$ particles in $2 n_F$
spin orbitals. Solving the Hamiltonian in the Hilbert space of the impurity
corresponds to a Fock space calculation in the fragment subspace. The bath can
be considered as a reservoir of electrons or holes.

Having solved the impurity Hamiltonian, the energy density
(energy per fragment) is subsequently computed as
\begin{align}
\label{EnExpr}
E = \sum_{f e} \tilde{h}_{f e} \gamma_{e f}
+ \frac{1}{4} \sum_{f e e^{\prime} e^{\prime \prime}}
\tilde{V}_{f e e^{\prime} e^{\prime \prime}} \Gamma_{e^{\prime} e^{\prime \prime} f e} ,
\end{align}
with $\gamma_{e e^\prime} = \langle e^{\prime\dagger} e \rangle$ and
$\Gamma_{e e^\prime e^{\prime\prime} e^{\prime \prime \prime}} =
\langle e^\dagger e^{\prime\dagger} e^{\prime \prime \prime} e^{\prime \prime} \rangle$
being one- and two-particle density matrices of the
impurity wavefunction. Again, index $f$($e$) denotes fragment (fragment and bath)
single-particle states.

We note that the DET energy is not an expectation
value of the true Hamiltonian with an N-particle wavefunction. It is therefore not
an upper bound of the true ground state energy.

\subsection{Schmidt decomposition for periodic systems}

Density embedding calculations on a truly infinite system require a suitable
single-particle basis associated with a fragment. In the present work, we
employ the maximally localized Wannier functions \cite{RevModPhys.84.1419,WannWC1,WannWC2,Zicovich-Wilson}
obtained by localization of canonical mean-field crystalline orbitals. In
other words, we form a unitary transformation of mean-field basis
$|\psi_{n\vec{k}}\rangle$, where $\vec{k}$ labels irreducible representation of
the translational group \cite{pisani1996quantum,CPL.289.611.1998} and $n$ is a band
index. This yields an orthonormal set $|F_{i \vec{G}}\rangle$, where
$i$ labels a basis in given cell $\vec{G}$. The orthonormality condition
reads, \cite{RevModPhys.84.1419}
\begin{align}
\langle F_{i \vec{G}} | F_{j \vec{G}^{\prime}} \rangle = \delta_{ij} \delta_{\vec{G}\vec{G}^\prime} .
\end{align}

A few comments are called for at this point. Firstly, during the localization
process, we must allow for mixing of hole and particle states.  Therefore,
there is no need for localizing the particle (unoccupied) orbitals by
themselves. In our numerical approach, we did not encounter serious
difficulties converging the Wannier basis required by the present formalism.
Secondly, as we explain in more details in Appendix \ref{BandTrun}, one may
desire to truncate the space treated in the impurity Hamiltonian only to
levels around the Fermi energy. This is accomplished simply by choosing a
subset of energy bands for the localization. In other words, only a limited set
of the highest valence bands and the lowest conduction bands may be employed
while forming $|F_{i\vec{G}}\rangle$ bases. As the number of bands used during
the localization process is equal to the number of fragment states per unit
cell, a suitable truncation criterion may be used to further limit the
single-particle Hilbert space of the impurity Hamiltonian without sacrificing
relevant physics.

Analogously, one needs to perform the localization of the hole states, yielding
$|\phi_{p\vec{G}}\rangle$, which constitutes the $p^{\textrm{th}}$ hole state
associated with cell $\vec{G}$. Whenever a truncation of the conduction band
occurs while forming the fragment states, the same truncation should be done
during the formation of the hole states.

Now, one is in a position to compute the overlap matrix of Eq. \ref{MMat},
\begin{align}
\label{MMatPBC}
\mathbb{M}{}_{p q}^{\vec{G}\vec{G}^\prime} = \langle \phi_{q\vec{G}^\prime} | \hat{P}_F | \phi_{p\vec{G}} \rangle =
\sum_{i} \langle \phi_{q\vec{G}^\prime} | F_{i\vec{0}} \rangle \langle F_{i\vec{0}} | \phi_{p\vec{G}} \rangle ,
\end{align}
needed to define the fragment and bath states of the impurity basis. In the
above formula, the summation over the fragment states is limited to the reference cell
$\vec{0}$.  If embedding of more than one cell is needed, the summation over
the entire embedded cluster must be performed. With the aid of the above
matrix, fragment and bath states are formed according to Eq.
\ref{ConFrag} keeping in mind that index $p$ in those equations
includes a cell coordinate.

At this point we would like to note that the localization of the hole states
and the local nature of the fragment single-particle basis allows for effective
truncation of the formally infinite summation over the entire crystal. Indeed, one
may limit the summation over cells when $\langle F_{i\vec{0}} | \phi_{p\vec{G}}
\rangle \to 0$ as $|\vec{G}|$ increases. The localization of the hole states
therefore dictates a natural length scale one has to consider during DET
calculations.

\subsection{Definition of the impurity Hamiltonian}

Having established a formalism for constructing the embedding basis, let us
turn our attention to the definition of the impurity Hamiltonian.
A Hamiltonian for a realistic crystalline material can be written as
\begin{align}
\hat{H} = E_{NN} + \hat{V}_{Ne} + \hat{V}_{ee} + \hat{T} = H_0 + \hat{h} + \hat{V},
\end{align}
where $E_{NN}$ is the nuclear repulsion energy, $\hat{V}_{Ne}$
and $\hat{V}_{ee}$ are the electrostatic electron-nucleus and electron-electron
interactions, respectively; $\hat{T}$ is the kinetic energy operator. Those terms
can be then arranged into constant ($H_0$) and one- and two-body Hamiltonians
($\hat{h}$ and $\hat{V}$, respectively). As described in the literature,
\cite{PhysRevB.61.16440,pisani1996quantum,JCP-105-10983-1996,PhysRevB.28.5781}
the summation of an infinite number of electrostatic terms has to be handled
with care in order to avoid divergences and loss of accuracy. Analogous problems
may arise while projecting the Hamiltonian onto the embedding basis.  In order to
deal with such complications, we propose to first recast the Hamiltonian into
second-quantized form with the aid of the mean-field Fock matrix
\begin{align}
F_{\mu \nu} = h_{\mu \nu} + \sum_{\lambda \sigma}
V_{\mu \lambda \nu \sigma} \gamma_{\sigma \lambda} ,
\end{align}
as
\begin{align}
\hat{H} & = E_0 - \sum_{\mu \nu} F{}_{\mu\nu} \gamma_{\nu\mu}
             + \frac{1}{2} \sum_{\mu \nu \lambda \sigma}
             V_{\mu \lambda \nu \sigma} \gamma_{\sigma \lambda} \gamma_{\nu\mu}    \nonumber \\
        &    + \sum_{\mu \nu}
             \Big(F_{\mu\nu} - V_{\mu \lambda \nu \sigma} \gamma_{\sigma \lambda} \Big) c{}^\dagger_\nu c_\nu
         + \frac{1}{4} \sum_{\mu \nu \lambda \sigma}
             V_{\mu \nu \lambda \sigma} c{}^\dagger_\mu c{}^\dagger_\nu c_\sigma c_\lambda
\end{align}
In the expression above, $E_0 = E N$ where $E$ is the mean-field energy per
unit cell and $N$ is the number of cells. Again, the individual terms in
the summation are not necessarily convergent. For example, the constant
$\frac{1}{2} \sum_{\mu \nu \lambda \sigma} V_{\mu \lambda \nu \sigma} \gamma_{\sigma \lambda} \gamma_{\nu\mu}$
describing the electron-electron interaction energy is divergent and has no
meaningful thermodynamic limit, if evaluated separately. Similarly,
$\sum_{\lambda \sigma} V_{\mu \lambda \nu \sigma} \gamma_{\sigma \lambda}$,
which contributes to the one-body Hamiltonian above, gives rise to divergent
matrix elements. For these reasons, we propose to express all quantities in the
embedding basis, before the summation is performed.  In other words, we
separately project the mean-field potential, the density matrix and the two-body
interaction onto the embedding basis, i.e. $F_{\mu \nu} \to \tilde{F}_{e
e^\prime}$,
$\gamma_{\mu \nu} \to \tilde{\gamma}_{e e^\prime}$ and
$V_{\mu \nu \lambda \sigma} \to \tilde{V}_{e e^\prime e^{\prime \prime} e^{\prime \prime \prime}}$.
While the two-body interaction is projected without any modifications, let us explicitly write the
one-body part of the impurity Hamiltonian, $\tilde{h}$, and the constant term, $\tilde{E_0}$,
\begin{align}
\label{ReDefOne}
\tilde{h}_{e e^\prime} &=  \tilde{F}_{e e^\prime} -
\sum_{e^{\prime \prime} e^{\prime \prime \prime}} \tilde{V}_{e e^{\prime \prime} e^\prime e^{\prime \prime \prime}}
\tilde{\gamma}_{e^{\prime \prime \prime} e^{\prime \prime}}
\end{align}
\begin{align}
\label{ReDefZero}
\tilde{E_0} &= E N_F - \sum_{f e} \Big( \tilde{F}_{f e} - \frac{1}{2}\sum_{e^\prime e^{\prime \prime}}
\tilde{V}_{f e^\prime e e^{\prime \prime}} \tilde{\gamma}_{e^{\prime \prime} e^\prime} \Big)\tilde{\gamma}_{e f} ,
\end{align}
where again, $E$ denotes the mean-field energy per unit cell, whereas $N_F$ is the number of unit cells
in the fragment. As the reader may readily notice, the summation restriction to the fragment
basis only has been imposed on the constant term above. The reason for such truncation
shall become clear soon.

The construction above constitutes an approximate way of projecting the Hamiltonian.
Let us therefore discuss the physical motivation behind it.
As shown in Ref. \onlinecite{PhysRevB.89.035140} and expanded upon
in the Appendix below, the mean-field one-particle density matrix and mean-field Fock
matrix commute with each other after projection onto the embedding basis,
i.e., $[\tilde{\gamma},\tilde{F}]=0$. Moreover, $\tilde{\gamma}$ is idempotent.
Inserting $\tilde{\gamma}$ as an initial guess for the impurity Hamiltonian as
define above yields a Fock matrix that is equal to the Fock matrix of the whole system
projected onto the embedding basis,
\begin{align}
F{}^{\textrm{imp}}_{e e^\prime} = \tilde{h}_{e e^\prime} + \sum_{e^{\prime \prime} e^{\prime \prime \prime}}
\tilde{V}_{e e^{\prime \prime} e^\prime e^{\prime \prime \prime}}
\tilde{\gamma}_{e^{\prime \prime \prime} e^{\prime \prime}} = \tilde{F}_{e e^\prime} .
\end{align}
The mean-field solution of the impurity problem is therefore the crystal density matrix
in the embedding basis. Furthermore, computing the energy according to Eq.
\ref{EnExpr} (with a constant term defined by Eq. \ref{ReDefZero}) reveals that
the mean-field energy of the fragment is just the energy per unit cell
multiplied by the number of cells taken as fragment constituents. In the above,
we have set the effective potential, present in Eq. \ref{FullHam} to zero. We
conclude that the current definition of the impurity problem ensures exactness
of mean-field in mean-field embedding. Furthermore, the solution
corresponds to a vanishing effective potential $v$. We would like to stress that
the exactness of the mean-field in mean-field embedding has been
numerically demonstrated for DMET in Ref. \onlinecite{JCTC.9.1428}.

Finally, let us note that for nontrivial calculations, that is when
a correlated theory is used as an impurity solver, the effective
potential has to be optimized and included in the impurity Hamiltonian as well
as the full crystal Hamiltonian. In the present work, the diagonal ansatz for
the effective potential allows us to eliminate these terms from the mean-field
Fock matrix of the crystal as it cannot change the mean-field solution.
Therefore, the construction of the embedding basis and the impurity Hamiltonian
is performed only once during the calculation.  The value of the effective
potential is determined in the embedding basis only.

\section{Computational details}

The construction of Wannier functions has been implemented in the
Gaussian Developement Version \cite{gdv} that has also been used to perform
the periodic Hartree-Fock calculations. The crystalline orbital localization
has been performed by adapting the scheme the of Ref. \onlinecite{Zicovich-Wilson},
where the Boys localization is replaced by the Pipek-Mezey localization
\cite{WannWC2} with the L{\"o}wdin population. \cite{WannWC1}
For 1D systems, we have used a $\vec{k}$-point
mesh of at least 400 points; for 2D and 3D, 4000 and 70000 $\vec{k}$-points
have been used, respectively.
The hermitized density matrices for coupled cluster with double (CCD) and single and double (CCSD)
excitations were obtained using the linear response formalism.
\cite{JCP.1995.103.3561,JCP.1987.87.5361}

The most diffuse basis functions of the 6-31G basis \cite{JCP.1972.56.2257,JCP.1975.62.2921}
were changed to 0.35, 0.30, and 0.20
for the carbon, nitrogen and boron atoms, respectively.

In all calculations, eigenvalue thresholds of the Schmidt decomposition
for retaining the bath states was set to 10$^{-6}$. The fragment states corresponding to
eigenvalues that were closer to 0 or 1 than this threshold were constructed according to
the formalism outlined in Appendix \ref{LinDep}.

The number of cells used in the Schmidt decomposition has been decided
by a commutation criterion between mean-field Fock and density matrices
after projection onto the embedding basis,
$\sum_{ee^\prime}|(\tilde{F}\tilde{\gamma})_{ee^\prime} - (\tilde{\gamma}\tilde{F})_{ee^\prime}|$.
The values of this norm are reported for the calculations
in the subsequent section.

\section{Results and discussion}

\subsection{1D carbon systems}

In this section, we asses the performance of DET on
three carbon polymers, polyyne (C$\equiv$C)$_\infty$,
polyacetylene (CH$=$CH)$_\infty$, and polyethylene
(CH$_2$$-$CH$_2$)$_\infty$. In the present work, we adopt the
geometries from Ref. \onlinecite{JCP.120.2004.2581}.
The geometrical parameters are $r_{\mathrm{C}\equiv\mathrm{C}} = 1.263 \AA$,
$r_{\mathrm{C}-\mathrm{C}} = 1.132 \AA$ for polyyne, $r_{\mathrm{C}=\mathrm{C}} = 1.369 \AA$,
$r_{\mathrm{C}-\mathrm{C}} = 1.426 \AA$, $r_{\mathrm{C}-\mathrm{H}} = 1.091 \AA$,
$\angle_{\mathrm{C}=\mathrm{C}-\mathrm{C}} = 124.5^\circ$, $\angle_{\mathrm{C}=\mathrm{C}-\mathrm{H}} = 118.3^\circ$
for polyacetylene, and $r_{\mathrm{C}-\mathrm{C}} = 1.534 \AA$, $r_{\mathrm{C}-\mathrm{H}} = 1.100 \AA$,
$\angle_{\mathrm{C}-\mathrm{C}-\mathrm{C}} = 113.7^\circ$, $\angle_{\mathrm{H}-\mathrm{C}-\mathrm{H}} = 106.1^\circ$
for polyethylene. In order to gain better insight into the performance of the DET
approximation, we have additionally deformed the above systems by keeping all
variables, apart from the carbon-carbon bonds, fixed, while scaling the
carbon-carbon bonds uniformly with a parameter $\alpha$. In all calculations,
the $1s$ orbitals of carbon were eliminated from consideration in DET and
coupled cluster calculations. DET($n$) denotes calculations with $n$ unit cells
used as a fragment.

\begin{figure}[t!]
\begin{center}
{\resizebox{\imsize}{!}{\includegraphics{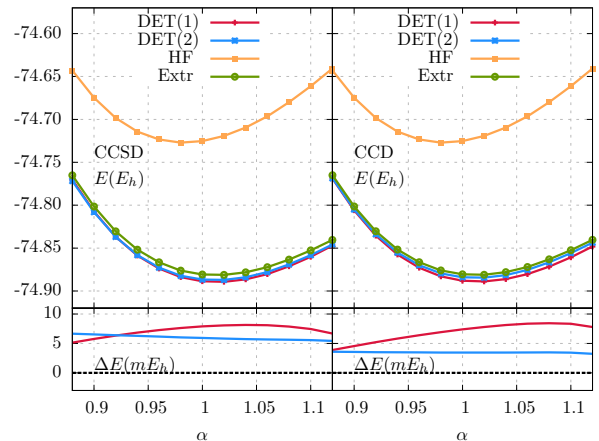}}}
\end{center}
\caption{Energy per unit cell profile for polyyne (C$\equiv$C)$_\infty$ with
respect to uniform deformation (see text for details) with STO-3G basis.  The
results of DET(1) and DET(2) with CCSD (left panel) and CCD (right panel) as
impurity solver calculations are compared to CCSD (left panel) and CCD (right
panel) oligomeric extrapolation (Extr) and Hartree-Fock (HF). For clarity the
difference between DET and extrapolated data is displayed in the bottom panel.
}
\label{Fig.C2.STO3G}
\end{figure}

\begin{figure}[t!]
\begin{center}
{\resizebox{\imsize}{!}{\includegraphics{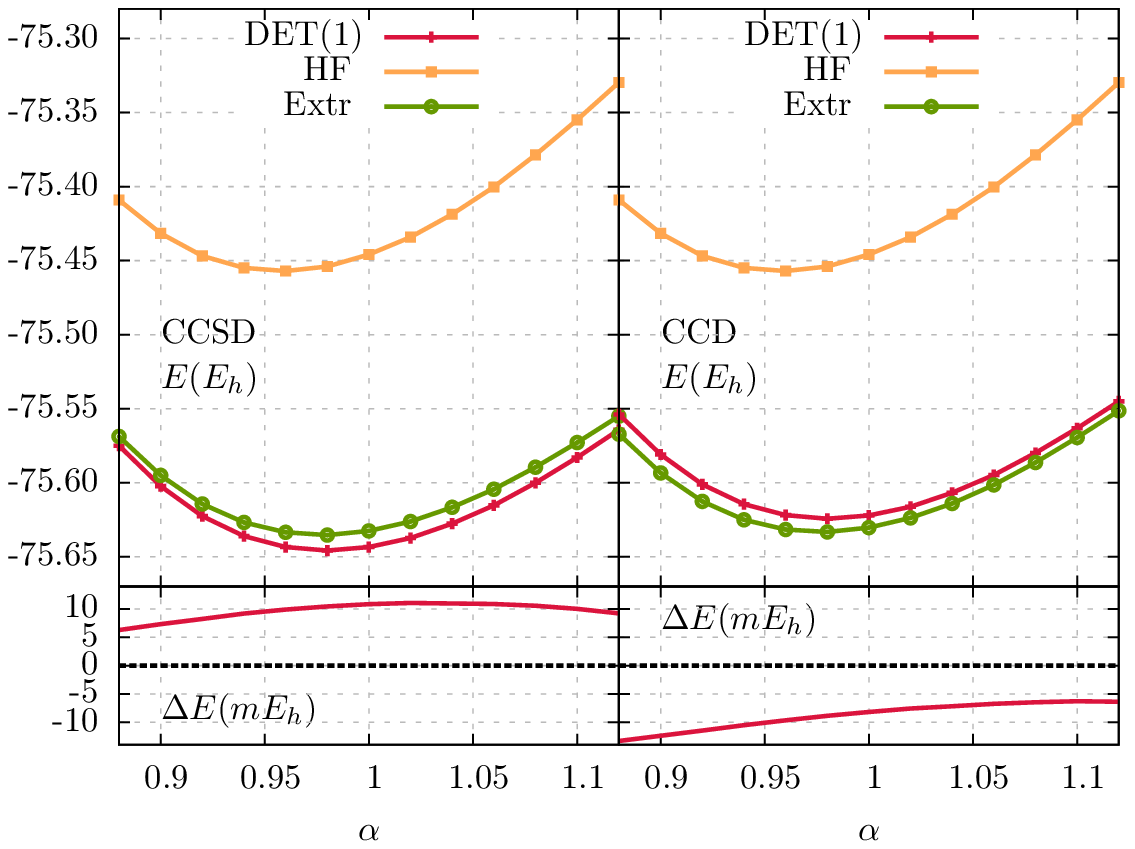}}}
\end{center}
\caption{Same as Fig.\ref{Fig.C2.STO3G} with 6-31G basis.}
\label{Fig.C2.631G}
\end{figure}

Let us begin the discussion with the most challenging system for the embedding
calculation, polyyne. In this case, one expects that the correlation energy
contribution to the unit cell would have the slowest decay. \cite{JCP.120.2004.2581}
The results are shown in Fig. \ref{Fig.C2.STO3G} and Fig. \ref{Fig.C2.631G}.
The extrapolated CCD and CCSD results were obtained according to
\begin{align}
E_{\textrm{Extr}}(n) = E_{HF} + E_{\textrm{corr}}(n) - E_{\textrm{corr}}(n-1)
\label{Extrapolation}
\end{align}
where $E_{HF}$ is HF energy per unit cell of infinite system and
$E_{\textrm{corr}}(n)$ is the correlation energy of the $n$-unit oligomer with a
hydrogen atom as the terminal group.  For the case of STO-3G basis,
extrapolations for $n=8$ and $n=7$ differ by no more than 0.1 $mE_h$; in the
case of the 6-31G basis, we have used $n=9$ which differs from $n=8$ by at most
0.2 $mE_h$. We deem these results sufficiently converged for the purpose of the
presented figures. Our extrapolated CCSD correlation energy value for the STO-3G basis
and $\alpha=1$, of -155.45 $mE_h$ agrees well with -155.53 $mE_h$ reported in Ref.
\onlinecite{JCP.120.2004.2581}.  In the DET calculations,
the number of cells used for the Schmidt decomposition guaranteed that the norm of the
commutator of full-system Fock and density matrices projected onto the embedding basis
to be at most $3\times 10^{-6}$.

As is clear from Fig. \ref{Fig.C2.STO3G}, the DET calculations with a single
unit cell chosen as a fragment agree well with the extrapolated thermodynamic
limit values both for CCSD and CCD as the impurity solver.  The maximum
discrepancy is below 10 $mE_h$; this translates to an energy difference on the
level of $5\%$. Investigating the shape of the energy profile as a function of
the uniform stretching parameter $\alpha$, we find the overall agreement
satisfactory. The inclusion of electron correlation clearly favors a more
stretched configuration. Apparently, DET calculations appropriately capture
this trend. One may also notice that including two unit cells as the fragment
yields results that are closer to the extrapolated CCSD and CCD results.

Increasing the size of the basis set to 6-31G does not lead to a deterioration
of the DET results. As is clear from Fig. \ref{Fig.C2.631G}, the single-cell
DET calculations are again in good agreement with the extrapolated values. The
absolute difference does increase slightly but so does the correlation energy.
The overall shape of the energy profile is well reproduced by DET. The bonds
elongation caused by correlation is well captured.

Let us comment on the size of the impurity problem for polyyne. In
the case of the STO-3G basis, there are 8 fragment and 8 bath orbitals for the
single cell case, with the impurity bearing 16 electrons.  This illustrates how
effective the present embedding scheme is in truncating the size of the
single-particle Hilbert space of the problem.

\begin{figure}[t!]
\begin{center}
{\resizebox{\imsize}{!}{\includegraphics{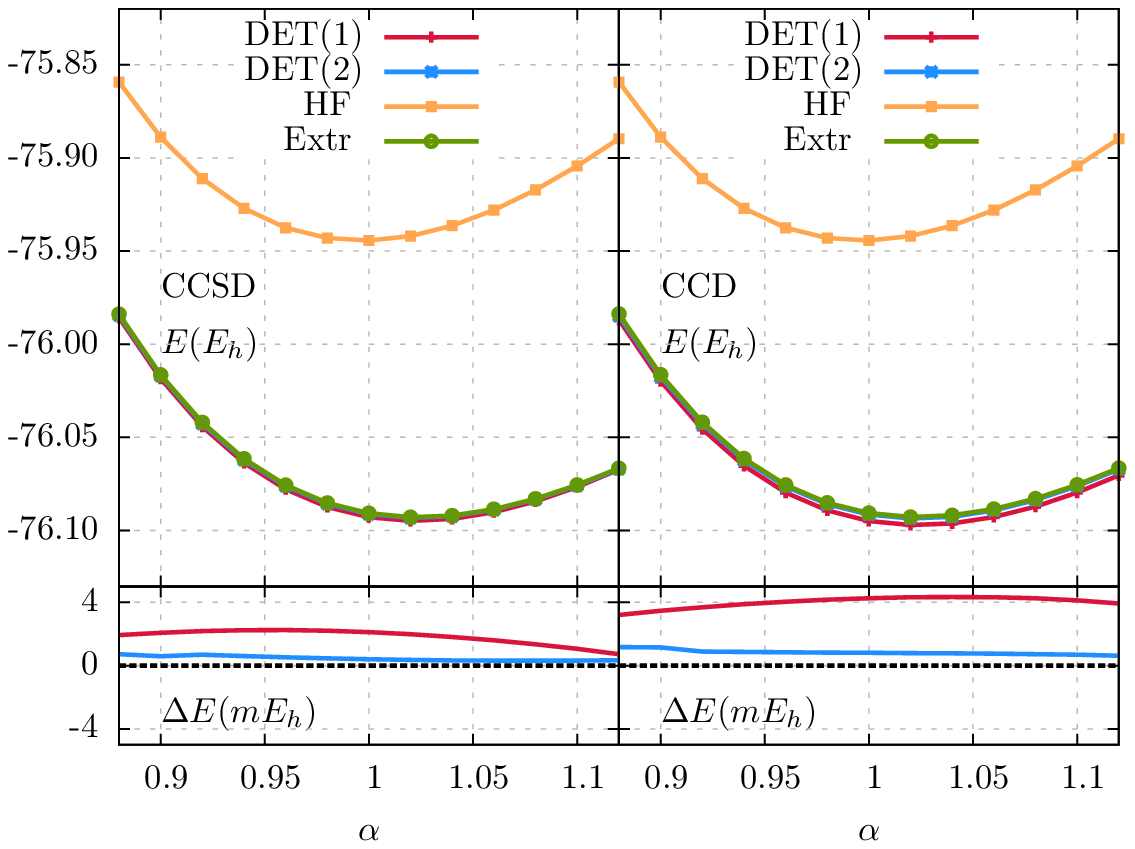}}}
\end{center}
\caption{Same as Fig.\ref{Fig.C2.STO3G} for polyacetylene.}
\label{Fig.CH.STO3G}
\end{figure}

\begin{figure}[t!]
\begin{center}
{\resizebox{\imsize}{!}{\includegraphics{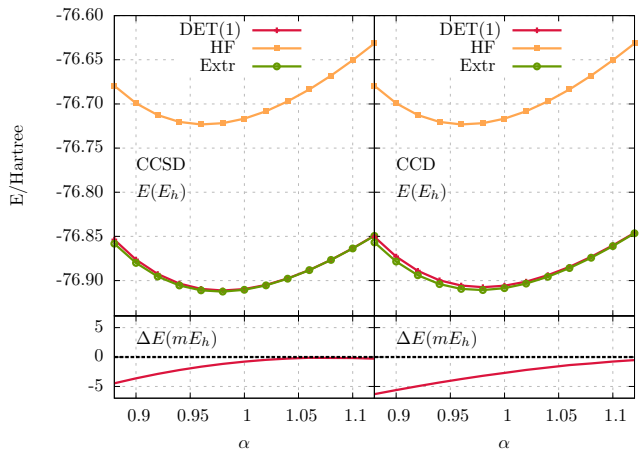}}}
\end{center}
\caption{Same as Fig.\ref{Fig.C2.STO3G} for polyacetylene with the
6-31G basis.} \label{Fig.CH.631G}
\end{figure}

The next polymeric system under investigation is
polyacetylene. For this example, the extrapolated CCSD and CCD correlation
energy from 8 and 7 cells differed by less than 0.1 $mE_h$ for both
STO-3G and 6-31G bases. The value of $E_{\textrm{corr}}$ for CCSD with $\alpha=1$,
-146.4 $mE_h$, coincides with the one reported in Ref. \onlinecite{JCP.120.2004.2581}.
The number of cells used in the Schmidt decomposition guaranteed that the norm
of the commutator between the mean-field density and Fock matrices in the embedding basis
is below $10^{-6}$.

Analogously to polyyne, one notices in Fig. \ref{Fig.CH.STO3G} and
Fig. \ref{Fig.CH.631G} that correlation favors a more elongated
carbon-carbon bond. Both DET and extrapolated oligomeric
results agree quantitatively. For the STO-3G basis, even the DET(1) calculation
yields results within 4 $mE_h$ from extrapolated values, a result that is greatly
improved by enlarging the embedded fragment to two cells. Regardless, the DET approximation
with both CCD and CCSD as impurity solvers yield results that
are rather parallel to the thermodynamic limit ones. With the increased size
of the basis set, the agreement remains satisfactory. Though the curvature of the
energy profile obtined with DET(1) deviates slightly from the extrapolated data,
especially for contracted systems ($\alpha \le 0.95$), the difference is not large.
Again, one has to keep in mind that DET calculations are done employing
a significantly truncated single-particle basis.
For the STO-3G basis, the impurity problem
with single cell models the infinite system with a Hamiltionian that describes
merely 20 electrons in 20 orbitals (10 fragment and 10 bath states).

\begin{figure}[t!]
\begin{center}
{\resizebox{\imsize}{!}{\includegraphics{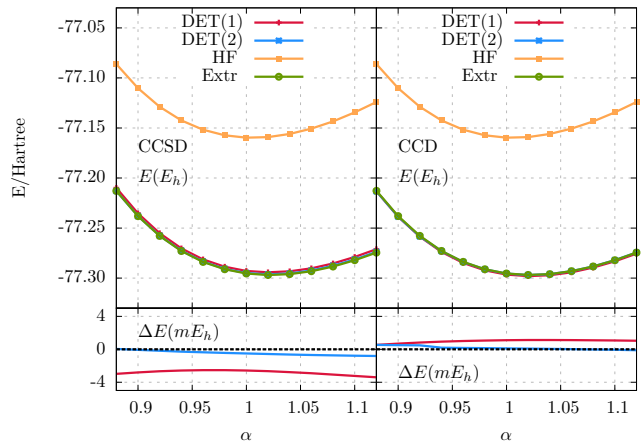}}}
\end{center}
\caption{Same as Fig.\ref{Fig.C2.STO3G} for polyethylene.}
\label{Fig.CH2.STO3G}
\end{figure}

\begin{figure}[t!]
\begin{center}
{\resizebox{\imsize}{!}{\includegraphics{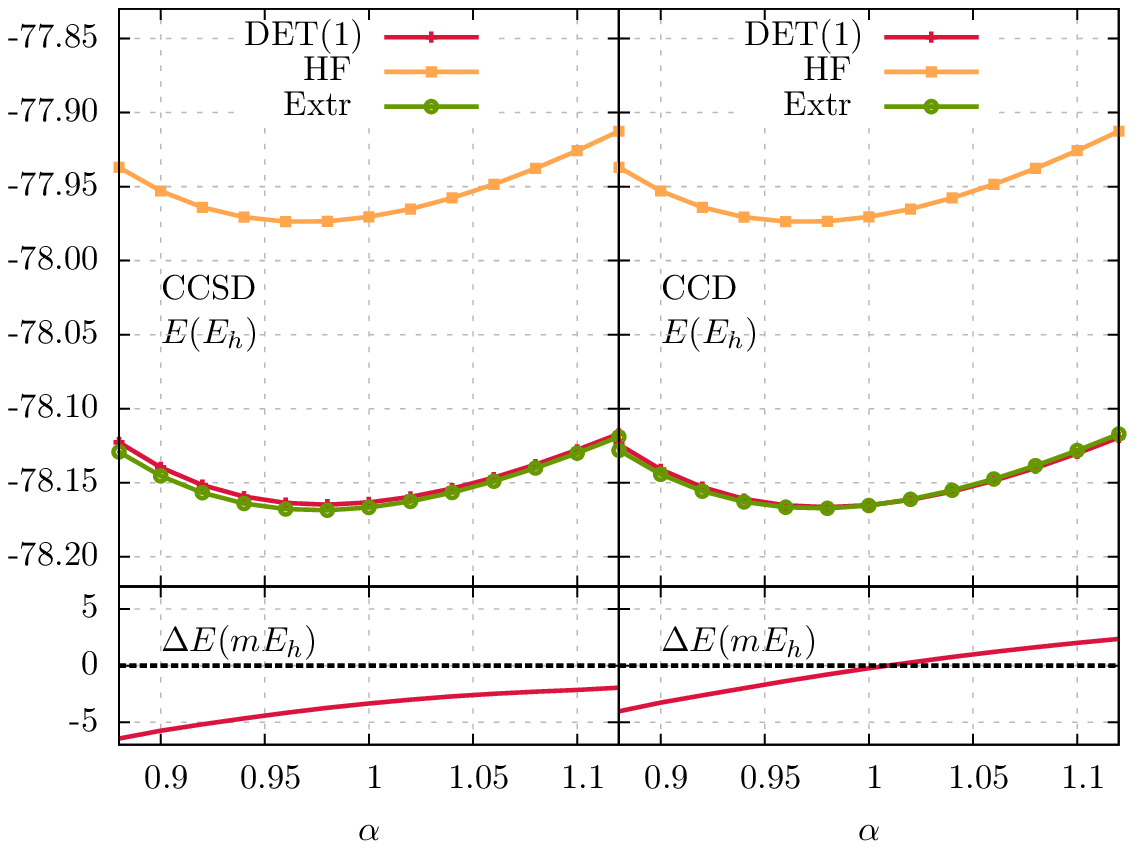}}}
\end{center}
\caption{Same as Fig.\ref{Fig.C2.STO3G} for polyethylene with the
6-31G basis.} \label{Fig.CH2.631G}
\end{figure}

The last 1D polymer studied is polyethylene. Just as in the
case of polyacetylene, the difference between the extrapolated CCD and CCSD
energy using 8 and 7 cells was well below 0.1 $mE_h$. For STO-3G and
$\alpha=1$, our extrapolated CCSD correlation energy per unit cell, -135.7
$mE_h$, coincides with the value reported by Hirata. \cite{JCP.120.2004.2581}
The number of cells included in the Schmidt decomposition guaranteed that the
norm of the commutator between the mean-field Fock and density matrices is below
$10^{-6}$ after projection onto embedding basis.

For polyethylene, the STO-3G DET(1) results (Fig. \ref{Fig.CH2.STO3G}) coincide
very well with the extrapolated oligomeric data. The small difference is almost
constant over the studied values of the $\alpha$ stretching parameter.
Increasing the embedded fragment to two cells brings the discrepancy almost to
zero. With the bigger basis, 6-31G, once again we observe very good overall
agreement between DET and extrapolated data. The maximum difference occurs for
the more contracted geometry.  Just as in all previous systems, one observes
stabilization of a more elongated structure due to correlation effects.  Again,
let us stress that, within the DET approximation, modeling the infinite system
with an impurity problem of 24 electrons in 24 orbitals (12 fragment states and
12 bath states), for the example of DET(1) STO-3G calculations, allows one to
obtain a high degree of agreement with the full periodic CCD and CCSD
calculations.  One should however keep in mind, that the physical
interpretation of the impurity problem differs from the true Hamiltonian. In
the former, the CC(S)D method is used to effectively perform a Fock space
calculation in the unit cell with the aid of an entangled bath. On the other
hand, for the full Hamiltonian one considers excitations of the electrons of
the entire periodic system.

\subsection{2D and 3D: boron nitride and diamond}

\begin{figure}[t!]
\begin{center}
{\resizebox{\imsize}{!}{\includegraphics{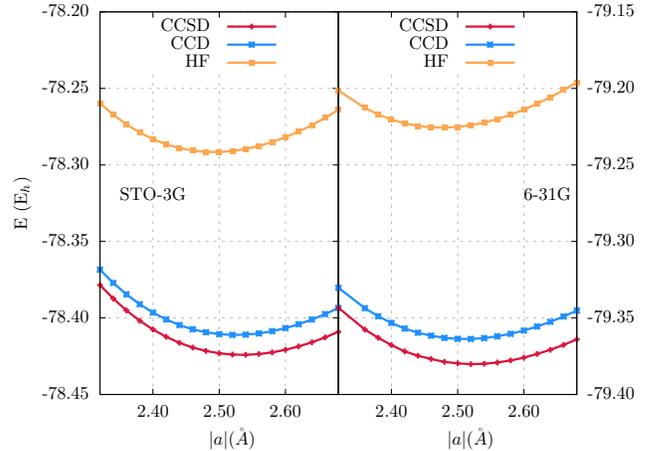}}}
\end{center}
\caption{Energy per unit cell as a function of lattice parameter of honeycomb
boron nitride lattice. The DET(1) calculations with CCSD and CCD as impurity solvers
are compared with HF for STO-3G (left) and 6-31G (right).
}
\label{Fig.BN}
\end{figure}

\begin{figure}[t!]
\begin{center}
{\resizebox{\imsize}{!}{\includegraphics{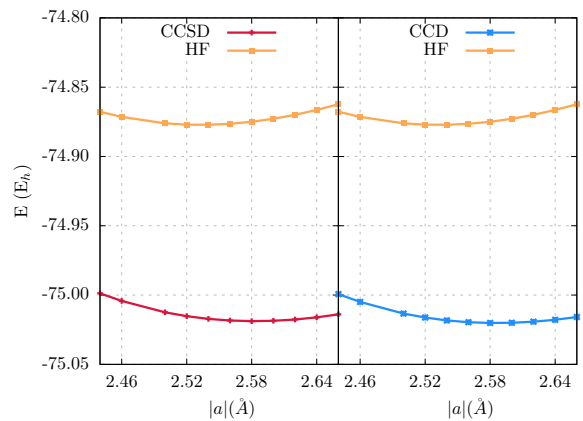}}}
\end{center}
\caption{Energy per unit cell as a function of lattice parameter (primitive unit cell)
for diamond. The DET(1) calculations with CCSD (left) and CCD (right) as impurity solvers
are compared with HF for the STO-3G basis.
}
\label{Fig.Diamond}
\end{figure}
In this section, we proceed to investigate prototypical 2D and 3D systems.
The 2D structure was obtained assuming infinitely separated boron nitride
sheets of hexagonal BN \cite{Science.2014.343.163,JCPCM.1997.1.1} yielding a graphene-like
honeycomb lattice. We performed a single unit embedding with CCSD and
CCD as an impurity solver. The number of cells included in the Schmidt decomposition
was chosen to provide the norm of the commutator of the mean-field and Fock matrices in the
embedding basis below 10$^{-6}$. The two lowest bands were excluded from consideration,
which corresponds to freezing 1$s$ orbitals of boron and nitrogen in the CC calculations.

In Fig \ref{Fig.BN}, we present the dependence of the
energy per unit cell with respect to the translation vector
defining the underlying honeycomb lattice. As the reader may readily notice,
our DET calculations predict noticeable impact from the inclusion of the single
excitation in the CC impurity solver. For both, the STO-3G and 6-31G bases,
the CCSD energy is consistently below the CCD energy by about 10-20 $mE_h$.
Nonetheless, the shape of the curves around the equilibrium point are
rather similar. The impact of the electron correlation clearly shifts
the position of the optimal structure towards longer translation
vectors as compared to mean-field calculations. We also note that
the lattice constant obtained with single cell DET embedding
is larger than that reported one for the hexagonal BN.
\cite{Science.2014.343.163,JAC.1994.2216.5} While this elongation is
due to the inclusion of only a single sheet,
lack of solid substrate or the deficiency of the employed
basis set is beyond the scope of the present work. However, we would like to
stress the key point of current work. Namely, the size of the
Hilbert space of impurity Hamiltonian is small and independent of the dimensionality
of the problem. The calculations for the single cell with STO-3G basis
required explicit correlated treatment of 16 electrons in 16 orbitals
while the 6-31G basis 32 electrons in 32 orbitals.

As a model 3D system, we have selected diamond. The single cell
embedding with the STO-3G basis is shown in Fig.
\ref{Fig.Diamond}. While in this example, the norm of the
commutator of Fock and density matrices in the embedding basis is
on the level of $3\times10^{-5}$, we have verified that even with
larger real space truncation, the DET correlation energy is stable
to 0.1 $mE_h$. As the reader may notice, CCD- and CCSD-based DET
calculations provide very similar descriptions. The equilibrium
geometry occurs for longer translational vector as compared to
Hartree-Fock calculations. We do not attempt a quantitative
discussion of the data presented, which would require larger
bases. However, we point out that the size of the impurity problem
involves 16 electrons in 16 orbitals (1$s$ orbitals were removed
from the impurity). We believe that this illustrates the key point
of the current work: the dimension of the impurity Hamiltonian is
independent of the dimensionality of the lattice.

\section{Conclusions}

In the present work, we have reported the first application of
density embedding theory for realistic periodic systems. We have
proposed a practical way of defining the impurity problem and an
extension of the Schmidt decomposition to infinite systems.
Practical aspects of calculation, including the problem of
disentangled bath states has been outlined and assessed. We
believe that this point is important for extending DET
calculations to larger basis sets and bigger fragments.

Our proposed formalism for realistic Hamiltonians has been
quantitatively assessed for several periodic systems with the aid
of coupled cluster theory as an impurity solver. The data
presented shows good agreement between the coupled cluster DET
calculations and the coupled cluster thermodynamic limit, even
when using a single cell as fragment. While more extended
benchmarks are certainly called for, the current tests are a
promising starting point. Indeed, employing more sophisticated
many-body techniques to tackle the impurity problem is an
interesting option especially as the size of the impurity
Hamiltonian does not depend on the dimensionality of the lattice.
Furthermore, as the impurity problem in DMET and DET is always
finite, application of accurate but not necessarily size-extensive
tools becomes feasible. Nonetheless, one should bear in mind that
DMET and DET provide a local, finite Hamiltonian solution which
may be interpreted as a Fock space calculation performed on the
fragment. As such, the philosophy of the calculation differs from
the Hilbert space approach to the full lattice.

Despite being a relatively recent model, density matrix embedding
theory and its simplification employed here, are accurate and
computationally feasible approaches to deal with the numerous
electronic degrees of freedom of large systems. We believe that
the results presented here support confidence in the predictive
power of this approximation.

\section{Acknowledgements}
I. W. B. would like to acknowledge Thomas M. Henderson for helpful
discussions. John Gomez and Jacob Wahlen-Strothman are thanked for
carefully reading the manuscript. This work was supported by
Department of Energy, Office of Basic Energy Sciences, Grant No.
DE-FG02-09ER16053. G. E. S. is a Welch Foundation Chair (C-0036).
\appendix*
\section{Handling band truncation and disentangled states in DET}

In this Appendix, we present a formalism for dealing with the Schmidt
decomposition of a Slater determinant for truncated particle and hole
states as well as cases where fragment and bath states become
disentangled. The discussion is similar to the one provided
in Ref. \onlinecite{PhysRevB.89.035140} but more general.

\subsection{Fragment states with truncated bands}
\label{BandTrun}
Let us start by specifying the notation. In the present work,
the crystaline orbitals, both in Bloch and Wannier representations
are normalized according to
\begin{align}
\langle \psi_{i\vec{k}} | \psi_{j\vec{k}^\prime} \rangle &= \delta_{ij} \delta_{\vec{k}\vec{k}^\prime} \nonumber \\
\langle \phi_{i\vec{G}} | \phi_{j\vec{G}^\prime} \rangle &= \delta_{ij} \delta_{\vec{G}\vec{G}^\prime} ,
\end{align}
and are related by the discrete Fourier relation \cite{RevModPhys.84.1419}
\begin{align}
| \phi_{i\vec{G}} \rangle &=  \frac{1}{\sqrt{N}} \sum_{\vec{k}} e^{-i \vec{k}\cdot\vec{G}}
\sum_{j} \mathbb{U}{}^{\vec{k}}_{ji} |\psi_{j\vec{k}} \rangle ,
\end{align}
where $N$ is the number of unit cells. The idempotent density matrix can then be expressed as
\begin{align}
\hat{\gamma} = \sum_{p\vec{k}} |\psi_{p\vec{k}}\rangle \langle \psi_{p\vec{k}} | =
\sum_{p\vec{G}} | \phi_{p\vec{G}} \rangle \langle \phi_{p\vec{G}} | =
\sum_{p^\prime} | \phi_{p^\prime} \rangle \langle \phi_{p^\prime} | ,
\end{align}
where in the second term, index $p$ denotes hole states at given $\vec{k}$ or labeled by
cell index $\vec{G}$, whereas in the last term
$p^\prime=(p\vec{G})$ denotes all particle states in all cells.

The orthonormal single-particle
basis $|F_{i\vec{G}}\rangle$ becomes
\begin{align}
|F_{i\vec{G}}\rangle = \frac{1}{\sqrt{N}} \sum_{\vec{k}} e^{-i \vec{k}\cdot\vec{G}}
\sum_{j}\mathbb{U}{}^{\vec{k}}_{ji}|\psi_{j\vec{k}}\rangle ,
\end{align}
with index $j$ running over the chosen subset of bands. Because the states $F_{i\vec{G}}$
are orthogonal to Wannier functions obtained by unitary transformation of Bloch
functions within the complementary subset of bands, such states have zero overlap
with $|F_{i\vec{G}}\rangle$. Therefore, for the sake of argument, one may include these
in the definition of the overlap matrix $\mathbb{M}$ (Eq. \ref{MMatPBC}). Such states
will simply have vanishing amplitude in eigenvectors corresponding to non-zero
eigenvalue. Therefore, in the rest of this Appendix, the summations over the
indices $p$ and $q$ in Eq. \ref{MMatPBC} are formally done over the whole valence band.
One just has to impose proper block-diagonal structure of $\mathbb{U}^{\vec{k}}$
while preparing states $|\phi_{p\vec{G}}\rangle$.
It now follows directly that the one particle density matrix in the
embedding basis has the blocked structure
\begin{align}
\tilde{\gamma} =
\begin{pmatrix}
\gamma^{\textrm{FF}} & \gamma^{\textrm{FB}} \\
\gamma^{\textrm{BF}} & \gamma^{\textrm{BB}}
\end{pmatrix}
=
\begin{pmatrix}
d & \sqrt{d(1-d)}  \\
\sqrt{d(1-d)} & 1-d
\end{pmatrix},
\end{align}
where $d$ is the diagonal matrix of non-zero eigenvalues of $\mathbb{M}$.
Finally, let us show the commutation relation between the density matrix and
the Fock matrix projected onto the embedding basis. To do so, we define
matrix $t = \tilde{F}\tilde{\gamma}$ and show it is Hermitian.
The fragment-fragment block reads,
\begin{align}
t{}^{\textrm{FF}}_{ij} &= \tilde{F}{}^{FF}_{ij} d_j + \tilde{F}{}^{FB}_{ij}\sqrt{d_j(1-d_j)}  \nonumber \\
&= \sqrt{d_i d_j}\sum_{pq} \mathbb{V}_{p i} \langle \psi_p | \hat{F} | \psi_q \rangle \mathbb{V}{}^\star_{q j}
= [t^{\textrm{FF}}_{ji}]^\star ,
\end{align}
where $\hat{F}$ is the crystal Fock operator. We have used the fact that a
vector $\hat{F}|\phi_q\rangle$ does not contain contributions from particle
states. Hence it can be written as $\sum_{r} |\phi_r\rangle \langle \phi_r |
\hat{F} | \phi_{q} \rangle$ with indices $q$ and $r$ being the hole states.
Similarly,
\begin{align}
t{}^{\textrm{BB}}_{ij} &= \sqrt{(1-d_i)(1-d_j)}\sum_{p q} \mathbb{V}_{pi} \langle \phi_p | \hat{F} | \phi_q \rangle
\mathbb{V}{}^\star_{qj} = [ t{}^{\textrm{BB}}_{ji} ]^\star  \nonumber \\
t{}^{\textrm{FB}}_{ij} &= \sqrt{d_i(1-d_j)} \sum_{pq} \mathbb{V}_{pi} \langle \phi_p | \hat{F} | \phi_q \rangle
\mathbb{V}{}^\star_{qj} = [ t^{\textrm{BF}}_{ji} ]^\star
\end{align}

\subsection{Handling disentangled states}
\label{LinDep}
In the following section we suggest a route for dealing with the eigenvalues of
$\mathbb{M}$ that are close to 1 or 0. As the reader may easily note,
whenever a situation like this occurs, one may face numerical problems
with the normalization of embedding basis. The trivial solution, i.e. removing
the couple of bath and fragment states that are disentangled, cannot be
easily done; one would eliminate possibly important degrees of freedom
in the fragment.

Let us consider that there exists a set of eigenvalues $d$ that are close
to 1. In such a case, we propose to remove the bath state and
retain a modified fragment state that takes the form
\begin{align}
\label{ZeroState}
|\tilde{f}_i\rangle = \sum_{p} \mathbb{V}{}^\star_{pi}  | \phi_p \rangle.
\end{align}
Due to the unitarity of $\mathbb{V}$, these states are orthonormal and orthogonal
to all other fragment and bath states. Moreover, the density matrix in
such a basis takes the form,
\begin{align}
\tilde{\gamma} =
\begin{pmatrix}
d              &  \sqrt{d(1-d)} & 0 \\
\sqrt{d(1-d)}  &  1 -d          & 0 \\
0              &   0            & 1 \\
\end{pmatrix},
\end{align}
where the last block is expressed in the basis $|\tilde{f}\rangle$. As is clear,
$\tilde{\gamma}$ remains idempotent. It is also straightforward to show that
the commutation relation between the one-body density matrix and
Fock matrix is preserved.

Let us now turn our attention to the situation when there exists
a set of eigenvalues $d$ that are close to 0. We propose
to construct an auxiliary matrix $\mathbb{N}$,
\begin{align}
\mathbb{N}_{kl} = \langle F_{l} | \big( \mathbb{I} - \sum_{p} |\phi_p\rangle \langle \phi_p | \Big) F_{k} \rangle .
\end{align}
This matrix admits eigendecomposition $\mathbb{N} = \mathbb{U} \lambda \mathbb{U}^\dagger$.
Let us show that for every eigenvalue $d_i$ of $\mathbb{M}$ different from 0,
$\mathbb{N}$ has eigenvalue $1-d_i$. We define a column vector
$\mathbb{U}^\prime_{ki} = \sum_{q} \langle F_k| \phi_q \rangle \mathbb{V}^\star_{qi}$
with a norm $\sum_{k} \mathbb{U}^{\prime\star}_{ki} \mathbb{U}^\prime_{ki} = d_i$.
Hence it is a non-trivial vector whenever $d_i$ is not an exact zero.
One may now explicitly verify that $\sum_{k}\mathbb{N}_{lk}\mathbb{U}^\prime_{ki} = (1-d_i) \mathbb{U}^\prime_{li}$

We show that one may replace a fragment state $|f_i\rangle$ with eigenvalue
close to 0, with the state
\begin{align}
|\bar{f}_i \rangle = \sum_{k} \frac{\mathbb{U}^{\star}_{ki}}{\sqrt{\lambda_i}} \big(\mathbb{I} - \sum_p
|\phi_p\rangle \langle \phi_p | \big) |F_k\rangle ,
\end{align}
with eigenvalue $\lambda_i = 1 - d_i$ and remove a bath state entangled with $|f_i\rangle$.
Let us now demonstrate that $\bar{f}_i$ corresponding to eigenvalue $\lambda_i$
is orthogonal to all fragment states $|f_j\rangle$ corresponding to $d_j$ not equal
to $1-\lambda_i$. Namely,
\begin{align}
\langle f_j | \bar{f}_i \rangle &=
\frac{1-d_j}{\sqrt{d_j \lambda_i}} \sum_{kp} \mathbb{V}_{pj}
\langle \phi_p | F_{k} \rangle \mathbb{U}{}^\star_{ki}  \nonumber \\
&=
\frac{\lambda_i}{\sqrt{d_j \lambda_i}} \sum_{kp} \mathbb{V}_{pj}
\langle \phi_p | F_{k} \rangle \mathbb{U}{}^\star_{ki}
\end{align}
must vanish whenever $\lambda_i \neq 1 - d_j$. Similarly, for the bath states
not associated with fragment $|f_j\rangle$,
\begin{align}
\langle b_j | \bar{f}_i \rangle = -\sqrt{\frac{d_j}{1-d_j}} \langle f_j | \bar{f}_i \rangle .
\end{align}
The orthogonality to states $|\tilde{f}_i\rangle$ (Eq. \ref{ZeroState}) also follows.
Finally, the density matrix in the embedding basis as define above takes the form
\begin{align}
\tilde{\gamma} =
\begin{pmatrix}
d              &  \sqrt{d(1-d)} & 0 & 0  \\
\sqrt{d(1-d)}  &  1 -d          & 0 & 0  \\
0              &   0            & 1 & 0  \\
0              &   0            & 0 & 0  \\
\end{pmatrix}.
\end{align}
One may also verify that the product of the Fock and density matrices in the embedding
basis remains Hermitian (one again uses the fact that
$\hat{F}|\phi_p\rangle = \sum_r|\phi_r\rangle\langle \phi_r | \hat{F} | \phi_p \rangle$
to show that $\langle \bar{f}_i | \hat{F} | \phi_p \rangle = 0$).

As is clear, the density matrix above is idempotent and traces to an integer
number of particles. Therefore, it dictates the number of electrons we include
in the impurity problem. However, as in the fragment space we replace
eigenvalues that differ from one and zero by a preset value, the total number of
particles in the fragment may deviate from the actual with an error
proportional to chosen threshold.

Let us finally note that in the derivation above we assumed that eigenvalues of fragment
states are arbitrarily small but nonzero. In practical calculations, we did not encounter
any problems while forming the fragment states from eigenvectors of $\mathbb{M}$ above the
preset threshold and filling the missing fragment states with the eigenvectors
of $\mathbb{N}$ to complete the set.
%
%
%
%
%
%

\end{document}